\documentclass[11pt,a4paper]{article}
\usepackage{amssymb,latexsym,a4wide}
\def\ZZ{{\mathbb Z}}

\def\CC{{\mathbb C}}
\def\RR{{\mathbb R}}
\def\proof{{\bf Proof}.\ }
\def\bull{\vrule height .9ex width .8ex depth -.1ex }

\newtheorem{formula}{}[section]
\newtheorem{proposition}[formula]{Proposition}
\newtheorem{definition}[formula]{Definition}
\newtheorem{corollary}[formula]{Corollary}
\newtheorem{remark}[formula]{Remark}
\newtheorem{lemma}[formula]{Lemma}
\newtheorem{theorem}[formula]{Theorem}

\def\thrm{\begin{theorem}}
\def\thrml#1{\begin{theorem}\label{#1}}
\def\ethrm{\end{theorem}}
\def\rmrk{\begin{remark}}
\def\rmrkl#1{\begin{remark}\label{#1}}
\def\ermrk{\end{remark}}
\def\dfntn{\begin{definition}}
\def\dfntnl#1{\begin{definition}\label{#1}}
\def\edfntn{\end{definition}}
\def\nmrt{\begin{enumerate}}
\def\enmrt{\end{enumerate}}

\def\qtn{\begin{equation}}
\def\qtnl#1{\begin{equation}\label{#1}}
\def\eqtn{\end{equation}}
\def\lmm{\begin{lemma}}
\def\lmml#1{\begin{lemma}\label{#1}}
\def\elmm{\end{lemma}}
\def\crllr{\begin{corollary}}
\def\crllrl#1{\begin{corollary}\label{#1}}
\def\ecrllr{\end{corollary}}
\begin{document}
\title{Probabilistic communication complexity over the reals}
\author{
Dima Grigoriev \\[-1pt]
\small CNRS, IRMAR, Universit\'e de Rennes \\[-3pt]
\small Beaulieu, 35042, Rennes, France\\[-3pt]
{\tt \small dmitry.grigoryev@univ-rennes1.fr}\\[-3pt]
\small http://perso.univ-rennes1.fr/dmitry.grigoryev
}
\date{}
\maketitle
\begin{abstract}
Deterministic and probabilistic communication protocols are introduced in which
parties can exchange the values of polynomials (rather than bits in the usual
setting). 
It is established a sharp
lower bound $2n$ on the communication complexity of recognizing the $2n$-dimensional
orthant, on the other hand the probabilistic communication complexity of its
recognizing does not exceed 4. A polyhedron and a union of hyperplanes are
constructed in $\RR^{2n}$ for which a lower bound $n/2$ on the probabilistic 
communication complexity of recognizing each  is proved. As a consequence
this bound holds also for the EMPTINESS and the KNAPSACK problems.
\end{abstract}

\section*{Introduction}

Communication complexity (see \cite{Yao}, a survey one can find in \cite{Nisan},
\cite{Lovasz}) in the usual (bit) setting counts the number of bit exchanges between
two (or more) parties 
who altogether compute a certain function
(one of the goals of the communication complexity was to provide a framework to analyze distributed computations
and to obtain lower bounds on other complexity ressources). 
In \cite{Ablayev} one can find the relations 
of the communication
complexity with the question of representing a function as a composition of functions of
a special form (this question stems from the  Hilbert's 13th problem). In \cite{Buhrman}
the communication complexity of quantum computations was studied.

In the present paper we introduce the model of communication protocols over real (or
complex) numbers when the parties exchange the values of polynomials. The variables of
polynomials are supposed to be partitioned in two groups: $X=\{X_1,\dots , X_{n_1}\},
Y=\{Y_1,\dots , Y_{n_2}\}$, the first party is able to calculate polynomials in $X$, the
second party in $Y$. It is worthwhile to mention that in \cite{krajicek} a different
(less restrictive) concept of a communication protocol was introduced in which the
parties can exchange arbitrary real numbers (rather than just values of a given family
of polynomials as in the present paper). After the present paper had been submitted the
paper \cite{Bl} has appeared in which a similar algebraic communication protocol was
introduced and several lower bounds on the algebraic communication complexity for computing
rational functions and recognizing algebraic varieties were established. Unlike \cite{Bl} 
we obtain lower bounds on {\it probabilistic} communication complexity and in addition,
for recognizing real {\it semi-algebraic} sets.

We note that parallel to the numerous customary (boolean or discrete) 
complexity classes one develops also their continuous (algebraic or semi-algebraic)
counterparts (see e.~g. \cite{Cucker}, \cite{Buergisser}). This paper
presents an attempt to introduce and study the probabilistic continuous
communication complexity.

For illustration of the results obtained in the present paper we consider the
KNAPSACK problem: whether for given sets $\{x_1,\dots,x_n\}$ and
$\{y_1,\dots,y_n\}$ there exist subsets $I_1,I_2\subseteq \{1,\dots, n\}$
such that $\sum_{i_1\in I_1}x_{i_1}+\sum_{i_2\in I_2}y_{i_2}=0$? There is
an evident {\it deterministic} communication protocol for the KNAPSACK problem
with the communication complexity $2n$ when two parties just yield 
$\{x_1,\dots,x_n\}$ and
$\{y_1,\dots,y_n\}$, respectively. In Section~\ref{last} we show a lower bound
$n/4$ in the complex case and $n/2$ in the real case on the {\it probabilistic}
communication complexity for the KNAPSACK problem. 

 In Section~\ref{Hessian} we define the {\it communication complexity of
computing a function} (polynomial for simplicity) and show a lower bound on it being the
rank of the matrix of its second derivatives, earlier this matrix in the frames
of communication complexity was employed in \cite{Abelson}. This slightly resembles the
lower bound on the bit communication complexity being the logarithm of the rank of the
communication matrix \cite{Yao}.

In Section~\ref{protocols} we describe the (deterministic) {\it communication protocols}
(respectively, {\it probabilistic communication protocols}) and relying on this we
define the (deterministic) {\it communication complexity of recognizing a set} 
(respectively, {\it probabilistic communication complexity}). As an application of the matrix of the second derivatives
 we establish a lower bound $n-3$ on a probabilistic communication complexity of
recognizing a constructible set in $\CC^{2n}$ whose Zariski closure contains the
hypersurface $\{f=X_1Y_1+\cdots +X_nY_n=0\}$. As a real counterpart we establish the
same bound $n-3$ for a semialgebraic set in $\RR^{2n}$ whose euclidean closure has
(full) $(2n-1)$-dimensional intersection with the hypersurface $\{f=0\}$.

In Section~\ref{orthant} we demonstrate a possible exponential gap between the deterministic
and probabilistic communication complexities. Namely, we prove a (sharp) lower bound
$n_1+n_2$ on the deterministic communication complexity of recognizing the orthant

$$\{(x_1,\dots , x_{n_1},y_1,\dots , y_{n_2})\in \RR^{n_1+n_2}: x_i>0, y_j>0,
1\leq i\leq n_1, 1\leq j\leq n_2\}.$$

\noindent
On the other hand, we show that the probabilistic communication complexity of recognizing
the orthant does not exceed 4.

In Section~\ref{protocols} the lower bound was established for a set which involves a
polynomial $f$ with a big communication complexity of its computation. In 
Section~\ref{last} we construct sets defined by linear contraints 
which nevertheless have big probabilistic communication complexity (clearly, any linear
function has the communication complexity of its computation at most 2). Namely, we
consider the polyhedron $\{X_i+Y_i>0, 1\leq i\leq n\} \subset \RR^{2n}$ and the
arrangement $\cup _{1\leq i,j\leq n} \{X_i+Y_j=0\} \subset \RR^{2n}$ and for each of both
prove a lower bound $n/2$ on the  probabilistic communication complexity of its
recognizing. For the complex arrangement $\cup _{1\leq i,j\leq n} \{X_i+Y_j=0\} \subset
\CC^{2n}$ we establish a lower bound $n/4$. As applications the obtained lower bounds
imply the same bounds for the EMPTINESS problem, i.~e whether $\{x_1,\dots , x_n\} \cap
\{y_1,\dots , y_n\} =\emptyset$, and for the KNAPSACK problem.

\section{Lower bound on the communication complexity of computing a function}\label{Hessian}

First we describe computational models for the communication complexity over
complex or real numbers. Let two families of variables $X=\{X_1,\dots, X_{n_1}\}$
and $Y=\{Y_1,\dots , Y_{n_2}\}$ be given. As usually in communication 
complexity studies, 
there are two parties. We assume that one party is able to calculate
polynomials $a_1(X),\dots, a_{r_1}(X)$ in $X$ and the second party is able to
calculate polynomials $b_1(Y),\dots, b_{r_2}(Y)$ in $Y$. Then the result is obtained
by means of calculating suitable polynomials $P_1( a_1(X),\dots, a_{r_1}(X), 
b_1(Y),\dots, b_{r_2}(Y)), \dots, P_N( a_1(X),\dots, a_{r_1}(X), 
b_1(Y),\dots, b_{r_2}(Y))$. The goal is to minimize $r_1+r_2$ viewed as a
measure of communication complexity.

We study the communication complexity of two problems: computing a 
polynomial $g(X,Y)$ and recognizing a subset $S$ in $(n_1+n_2)$-dimensional complex or
real space.

\begin{definition}\label{definition}
A polynomial $g(X,Y)$ has a {\it communication complexity} $c(g)$ less or equal to 
$r_1+r_2$ if $g=P( a_1(X),\dots, a_{r_1}(X), 
b_1(Y),\dots, b_{r_2}(Y))$ for appropriate polynomials $P,a_1,\dots,a_{r_1},b_1,
\dots, b_{r_2}$.
\end{definition}

Obviously, the communication complexity of $g$ does not exceed $n_1+n_2$.

By  $H(g)$ denote $n_1\times n_2$ matrix of the second derivatives
$({\partial^2 g \over \partial X_i \partial Y_j})$, by $H(P)$ denote $r_1\times r_2$
matrix $({\partial^2 P \over \partial a_{i_1} \partial b_{j_1}})$, by the Jacobian
$J(a_1,\dots , a_{r_1})$ denote $n_1\times r_1$ matrix of the first derivatives
$({\partial a_{i_1} \over \partial X_i})$, similar $J(b_1,\dots , b_{r_2})=
({\partial b_{j_1} \over \partial Y_j})$. Then we have

$$H(g)=J(a_1,\dots , a_{r_1}) H(P) (J(b_1,\dots , b_{r_2}))^T.$$

\begin{lemma}\label{lower}
(cf. \cite{Abelson}) In the notations of Definition~\ref{definition} we have 
$$c(g)\geq min\{r_1,r_2\} \geq rk (H(g)).$$
\end{lemma}

\begin{corollary}\label{inner}
$c(f=X_1Y_1+\cdots +X_nY_n)\geq n$
\end{corollary}

To deal in the sequel with communication protocols we need the following 
statement generalizing the latter corollary.

\begin{lemma}\label{divisor}
Let a polynomial $g$ be a multiple of $f$. Then $rk (H(g))\geq n-3$.
\end{lemma}

\proof  We write $g=f^mh$ where $f$ does not divide $h$ (evidently, $f$ is
absolutely irredicible when $n\geq 2$, we assume here that $n_1=n_2=n$). We have

$$H(g)=mf^{m-1}h\left({\partial^2 f \over \partial X_i \partial Y_j}\right) + f^m
\left({\partial^2 h \over \partial X_i \partial Y_j}\right)+$$

$$m(m-1)f^{m-2}h\left({\partial f \over \partial X_i}\right) \left({\partial f \over \partial Y_j}\right)+
mf^{m-1}\left({\partial f \over \partial X_i}\right) \left({\partial h \over \partial Y_j}\right)+
mf^{m-1}\left({\partial h \over \partial X_i}\right) \left({\partial f \over \partial Y_j}\right).$$

\noindent
Each of the latter three matrices has rank at most 1, so it suffices to verify that
the sum of the former two matrices divided by $f^{m-1}$ is non-singular, it equals 

$$M=mh\left({\partial^2 f \over \partial X_i \partial Y_j}\right) +
f\left({\partial^2 h \over \partial X_i \partial Y_j}\right)$$

\noindent
We have $\det (M) =(mh)^n+ff_1$ for a certain polynomial $f_1$, hence
$\det (M) \neq 0$. \bull

It would be interesting to clarify, whether one can majorate $c(g)$ via an
appropriate function in $rk(H(g))$?

\section{Probabilistic communication protocols}\label{protocols}

Now we define a {\it communication protocol} for recognizing a set $S$. We consider two 
cases: $S\subset \CC^{n_1+n_2}$ is a constructible set or $S\subset \RR^{n_1+n_2}$
is a semialgebraic set. A protocol is a rooted tree, and to its root an input
$(x,y)=(x_1,\dots , x_{n_1},y_1,\dots , y_{n_2})$ is attached. To every vertex $v$
of the tree (including the root, but excluding the leaves) either a certain polynomial
$a_v(X)$ or a polynomial $b_v(Y)$ is attached (so, it is calculated either by the 
first party or by the second party, respectively). To every vertex $v$ (of a depth
$r$) leads a unique path from the root, denote by $q_1(X,Y),\dots, q_r(X,Y)$ the
polynomials attached to the vertices $v_1,\dots, v_r=v$ along this path, thus for
every $1\leq k\leq r$ either $q_k(X,Y)=a_{v_k}(X)$ or $q_k(X,Y)=b_{v_k}(Y)$, respectively. 
In addition, to the vertex $v$ a family of {\it testing polynomials} $P_{v,1}(Q_1,\dots, Q_r),
\dots, P_{v,N_v}(Q_1,\dots, Q_r)$ is assigned. Similar to the usual decision trees
(see ~e.g. \cite{Meyer}, \cite{GKS}, \cite{G}) the protocol ramifies at $v$
according to the set of the signs $sgn(P_{v,1}(q_1(x,y),\dots, q_r(x,y))),\dots,
sgn(P_{v,N_v}(q_1(x,y),\dots, q_r(x,y)))$. Similar to decision trees in the
complex case the sign can attain two values: $=, \neq$, in the real case three
values: $=,<,>$. To every leaf a label either ``accept'' or ``reject'' is assigned
which provides an output of the protocol. To the protocol naturally corresponds a 
decision tree (without restrictions on the degrees of testing polynomials).
To any input $(x,y)$ corresponds a unique leaf of the protocol and a path
leading to this leaf, according to the signs of testing polynomials: the 
output assigned to the leaf is ``accept'' if and only if $(x,y)\in S$. 

The {\it communication complexity of the recognizing protocol} is defined as its
depth. We note that the communication complexity counts just the number of the
polynomials $a_{v_i}(X)$ or $b_{v_i}(Y)$, respectively, calculated (separately) by
each of both parties in several rounds along a path of the protocol and ignores the
(jointly) calculated polynomials $P_{v,1},\dots, P_{v,N_v}$.

Now we introduce {\it probabilistic communication protocols}. One can define it 
similar to probabilistic decision trees (cf. \cite{Meyer}, \cite{GKS}, \cite{GKMS},
\cite{G}) as
a finite family $C=\{C_i\}_i$ of  communication protocols $C_i$, 
chosen with
a certain probability $p_i\geq 0$, where $\sum_ip_i=1$. 
As for decision trees we require that a 
probabilistic communication protocol for any input returns a correct output with
the probability greater than 2/3 (we suppose that a certain continuous probabilistic
measure is fixed in the ambient space, ~e.g. one can take the Gaussian measure).
The maximal depth of communication protocols which constitute a 
probabilistic communication protocol is called the {\it probabilistic communication
complexity}.

First we consider probabilistic communication protocols over complex numbers.

\begin{proposition}\label{closure}
The probabilistic communication
complexity of an $(2n-1)$-dimensional constructible set $W\subset \CC^{2n}$
such that its Zariski closure $\overline W$ contains the hypersurface $U=\{f=X_1Y_1+\cdots
+X_nY_n=0\}$ is greater or equal to $n-3$.
\end{proposition}

\proof Let a probabilistic communication protocol $C$ recognize $W$. Among 
communication protocols which constitute $C$ there exists $C_0$ such that it gives
the correct outputs for at least of 1/3 of the points from $U$ and for at least of
1/3 of the points outside of $U$ (in fact, for the arguments below, instead of 1/3 any positive constant would
suffice).

Distinguish in the decision tree corresponding to $C_0$ a (unique) path along which
all the signs in the ramifications are $\neq$. Denote by $\{P_j(q_1(x,y),\dots,
q_r(x,y))\}_{1\leq j\leq N}$ the collection of all the testing polynomials along this
path, clearly $r$ does not exceed the communication
complexity of $C_0$. Denote $P=\prod_{1\leq j\leq N} P_j(q_1,\dots, q_r)$. Then the
inputs from the Zariski-open set $V=\{(x,y): P(x,y)\neq 0\}\subset \CC^{2n}$ follow 
this path in $C_0$.

Due to the choice of $C_0$ we conclude that $f$ divides $P$. Indeed, $C_0$
rejects all the points from a suitable (constructive) subset of $\CC^{2n}$ of the
dimension $2n$ because $C_0$ rejects a subset of a positive (namely, at least
1/3) measure, whence if $f$ did not divide $P$ then $C_0$ would reject all the
points of $U$ except for its certain (constructive) subset of the dimension at
most $2n-2$, but on the other hand, $C_0$ should accept a subset of a positive
measure (at least 1/3) from $U$.  Therefore, 
Lemma~\ref{divisor} and Lemma~\ref{lower} imply that $r\geq c(P)\geq rk (H(P))
\geq n-3$. \bull

For a semialgebraic set $S\subset \RR^{n_1+n_2}$ denote by $\partial (S) \subset
\RR^{n_1+n_2}$ its boundary, being a semialgebraic set as well. The following
proposition is a real counterpart of Proposition~\ref{closure}.

\begin{corollary}\label{boundary}
The probabilistic communication
complexity of a semialgebraic set $S$ such that $dim (\partial (S) \cap U)=2n-1$
is greater or equal to $n-3$.
\end{corollary}

\proof 
For any communication protocol $C_i$ from $C$ consider the product $P^{(C_i)}=
\prod_{1\leq j\leq N}P_j$ of all the testing polynomials from $C_i$ (cf. the proof of
Proposition~\ref{closure} where a similar product of the polynomials along a
particular path was taken). For any point $u \in \partial (S) \cap U$ there exists
$C_0$ such that $P^{(C_0)}(u)=0$, otherwise all the points from an appropriate ball
centered at $u$ would get the same output for all communication protocols $C_i$
from $C$ 
which would contradict the definition of the boundary. Hence there exists $C_0$ for
which $f$ divides $P^{(C_0)}$. Therefore, we complete the proof as at the end of 
Proposition~\ref{closure}. \bull

\section{Communication complexity of recognizing the orthant}\label{orthant}

Now we proceed to estimating the 
communication complexity of the orthant $T=\{(x_1,\dots, x_{n_1},y_1,\dots , 
y_{n_2})\in \RR^{n_1+n_2}: x_i>0, y_j>0, 1\leq i\leq n_1, 1\leq j\leq n_2\}$.
For this goal we use infinitesimals $\epsilon_1>\cdots >\epsilon_{n_1+n_2}>0$
(see ~e.g.\cite{GV}, \cite{GKS}, \cite{GKMS}, \cite{G}). Namely, denote by 
$\RR_i= \widetilde{\RR (\epsilon_1,\dots , \epsilon_i)}$ by recursion on $i$ the real
closure of the field $\RR (\epsilon_1,\dots , \epsilon_i)$, for the base of recursion
we put $\RR_0=\RR$. Then $\epsilon_{i+1}$ is transcendental over $\RR_i$ and for any
positive element $0<d\in \RR_i$ we have $0<\epsilon_{i+1}<d$. 

For a polynomial $g\in \RR [X_1,\dots, X_{n_1},Y_1,\dots, Y_{n_2}]$ denote by $lt(g)$
its {\it least term} with respect to the following (lexicographical) ordering: take the terms
with a minimal degree in $Y_{n_2}$, among them with a minimal degree in $Y_{n_2-1}$ and
so on. If $lt(g)=g_0 X_1^{i_1} \cdots X_{n_1}^{i_{n_1}}Y_1^{j_1}\cdots Y_{n_2}^{j_{n_2}}$
for a certain $g_0\in \RR$, we call $(i_1,\dots , i_{n_1}, j_1,\dots , j_{n_2})$ the
exponent vector of $lt(g)$. Take $e_1,\dots , e_{n_1+n_2} \in \{-1,1\}$, then we have
(cf. \cite{GKS}, \cite{GKMS}, \cite{G})

\begin{equation}\label{1}
sgn(g(e_1\epsilon_1,\dots, e_{n_1+n_2}\epsilon_{n_1+n_2}))=
sgn(lt(g)(e_1\epsilon_1,\dots, e_{n_1+n_2}\epsilon_{n_1+n_2}))
\end{equation} 

\begin{lemma}\label{exponent}
Let $g_1,\dots , g_s\in \RR [X_1,\dots , X_{n_1},Y_1,\dots , Y_{n_2}]$ and 
$P_1,\dots , P_N\in \RR [G_1,\dots , G_s]$. Then among the exponent vectors 
of the least terms of $P_1(g_1,\dots , g_s),\dots , P_N(g_1,\dots , g_s)$
there are at most $s$ linearly independent.
\end{lemma}

\proof We claim that if exponent vectors of any family of polynomials 
$h_1,\dots, h_t \in 
\RR [X_1,\dots , X_{n_1},Y_1,\dots , Y_{n_2}]$ are linearly independent then
 $h_1,\dots, h_t$ are algebraically independent over $\RR$. Indeed, denote 
the exponent vectors of $lt(h_1),\dots , lt(h_t)$ by $l_1,\dots , l_t$, respectively,
and denote by $L$ the $t\times (n_1+n_2)$ matrix with the rows $l_1,\dots , l_t$, then
for any polynomial $P=\sum_K p_KG^K \in \RR [G_1,\dots , G_t]$ the exponent of the 
least term of
$P(h_1,\dots , h_t)$ coincides with the least vector among the pairwise distinct 
vectors
$KL$ for all $K\in \ZZ^t$ such that $p_K \neq 0$. The proved claim entails the lemma 
immediately. \bull

\begin{theorem}\label{sharp}
The communication complexity of recognizing the orthant $T$ (as well
as its closure $\overline T$ in the euclidean topology) is greater
or equal to $n_1+n_2$.
\end{theorem}

\proof Let a communication protocol $C_0$ recognize $T$ (the 
arguing for $\overline T$ is similar). Using the Tarski's
transfer principle (see e.~g.
\cite{GV}, \cite{GKS}, \cite{GKMS}, \cite{G}) one can
extend the inputs of $C_0$ over the field $\RR_{n_1+n_2}$, then $C_0$ recognizes
the set $T^{(\RR_{n_1+n_2})}=\{(x_1,\dots, x_{n_1},y_1,\dots , 
y_{n_2})\in \RR_{n_1+n_2}^{n_1+n_2}: x_i>0, y_j>0, 1\leq i\leq n_1, 
1\leq j\leq n_2\}$. Take in $C_0$ the path which follows the input 
$(\epsilon_1,\dots, \epsilon_{n_1+n_2}) \in T^{(\RR_{n_1+n_2})}$. Let $r$ be the
length of this path and denote by $q_1(X,Y),\dots , q_r(X,Y)$ the polynomials
attached to the vertices along this path (we use the notations introduced in Section~\ref{protocols}  and
recall that every $q_i$ depends either on $X$ or on $Y$, although the latter
is not used in the proof of the Theorem, cf. Remark~\ref{difference} below). Let $P_1(q_1,\dots , q_r),
\dots , P_N (q_1,\dots , q_r)$ be all the testing polynomials along this path.

Lemma~\ref{exponent} implies that among the exponent vectors of 
$lt(P_1(q_1,\dots , q_r)), \dots, lt(P_N(q_1,\dots , q_r))$ 
there are at most $r$ linearly independent
$K_1,\dots , K_{r_0}, r_0\leq r$. Suppose that the theorem is wrong and $r<n_1+n_2$.
Pick a boolean vector $0\neq (m_1,\dots , m_{n_1+n_2}) \in (\ZZ /2\ZZ)^{n_1+n_2}$
orthogonal to all $K_i (mod \, 2), 1\leq i\leq r_0$. Then 

$$sgn(P_j (q_1,\dots , q_r) (\epsilon_1,\dots, \epsilon_{n_1+n_2})) =
sgn(P_j (q_1,\dots , q_r) ((-1)^{m_1}\epsilon_1,\dots, 
(-1)^{m_{n_1+n_2}}\epsilon_{n_1+n_2}))$$

\noindent
for $1\leq j\leq N$ (cf. the proof of lemma 1 \cite{GKS}). This means that the
output of $C_0$ is the same for the inputs $ (\epsilon_1,\dots, \epsilon_{n_1+n_2})$
and $((-1)^{m_1}\epsilon_1,\dots, 
(-1)^{m_{n_1+n_2}}\epsilon_{n_1+n_2})$. The obtained contradiction with the
supposition completes the proof of the theorem. \bull

\begin{remark}\label{difference}
The bound in Theorem~\ref{sharp} still holds if instead of communication
protocols one considers more general decision trees omitting the condition 
that each of the polynomials $q_1(X,Y),\dots,q_r(X,Y)$ depends either on $X$
or on $Y$. This strengthens slightly lemma 1 \cite{GKS} since here we consider
decision trees without a priori bound on fan-out of branching, unlike 
\cite{GKS} where the fan-out did not exceed 3.  
\end{remark}

\begin{remark}
Clearly, the communication complexity in the theorem equals  $n_1+n_2$.
\end{remark}

\begin{remark}
The probabilistic communication complexity of recognizing the closure $\overline T$
does not exceed $\log ^{O(1)} (n_1+n_2)$.
Indeed, the first party tests whether for an input $(x_1,\dots , x_{n_1},y_1,\dots ,
y_{n_2})$ the inequalities $x_1\geq 0, \dots , x_{n_1}\geq 0$ hold by means of a
probabilistic decision tree of the depth $\log ^{O(1)} n_1$ due to Theorem 1 
\cite{GKS}. The second party tests the inequalities $y_1\geq 0, \dots , 
y_{n_2}\geq 0$ by the same token.
\end{remark}

The latter remark demonstrates an exponential gap between the probabilistic and 
deterministic
communication complexities for recognizing the closure $\overline T$. The next
proposition provides even a bigger gap for $T$. 

\begin{proposition}
The probabilistic communication complexity of recognizing $T$ is at most 4.
\end{proposition}

\proof For an input $(x_1,\dots , x_{n_1},y_1,\dots ,
y_{n_2})$ consider the partition of the indices $\{1,\dots , n_1+n_2\}=I_0 \cup
I_{+}\cup I_{-}$ into the subsets for which the corresponding coordinates of the
input are zero, positive or negative, respectively. If $I_0\cup I_- \neq 
\emptyset$ then for a randomly chosen subset $I\subseteq \{1,\dots , n_1+n_2\}$
the probability of the event that $I\cap I_0 =\emptyset$ and that $|I\cap I_-|$ is
even is less or equal to 1/2. The latter statement is obvious when 
$I_0\neq \emptyset$, and when $I_0= \emptyset$ this probability equals to 1/2.

Therefore, when $(x_1,\dots , x_{n_1},y_1,\dots ,
y_{n_2}) \notin T$ and if one chooses randomly a product $\prod_{i_1\in I_1} x_{i_1}
 \prod_{i_2 \in I_2} y_{i_2}$ then this product is positive with the probability
less or equal to 1/2. Thus, the first party chooses randomly independently two subsets
$I^{(1)},I^{(2)} \subseteq \{1,\dots , n_1\}$ and calculates the products
$\prod_{i\in I^{(1)}} x_i$ and $\prod_{i\in I^{(2)}} x_i$ (in a similar way the second
party). If all 4 calculated products are positive then the output is ``accept'', otherwise
``reject''. \bull

\section{Lower bound on probabilistic communication complexity}\label{last}

Corollary~\ref{boundary} together with Lemma~\ref{divisor} show that if the
$(n_1+n_2-1)$-dimensional boundary of a semialgebraic set contains a ``facet''
with a great communication complexity of computing the polynomial which determines this
facet, then the probabilistic 
communication complexity of recognizing this set is great as well. Now we 
construct  a set (being a polyhedron) with a great probabilistic 
communication complexity (note that any facet of the polyhedron being determined by a linear
function, has a communication complexity at most 2).

Consider the polyhedron $S=\{(x_1,\dots , x_n,y_1,\dots ,
y_n) \in \RR^{2n}: x_i+y_i>0, 1\leq i\leq n\}$ and an {\it arrangement}
$R$ either real (i.~e. $\subset \RR^{2n}$) or complex (i.~e. $\subset \CC^{2n}$)
being a union of hyperplanes among which there appear $n$ hyperplanes
$\{X_i+Y_i=0\}, 1\leq i\leq n$.

\begin{theorem}\label{main}
The probabilistic communication complexity of recognizing over the reals the set $S$
or the set $R$ is greater than $n/2$.
\end{theorem}  

\proof Denote $Z_i=X_i+Y_i, \quad 1\leq i\leq n$. We consider the new coordinates 
$(X_1,\dots , X_n,Z_1,\dots , Z_n)$ in $\RR^{2n}$ and the point $u=(\epsilon_1,\dots , 
\epsilon_{2n})$. Let a probabilistic communication protocol $C$ recognize $S$
(respectively, $R$). Introduce $n$ points 
$u_i = (\epsilon_1,\dots , \epsilon_{n+i-1}, -\epsilon_{n+i}, \epsilon_{n+i+1},
\dots , \epsilon_{2n})$ (respectively,
$u_i^{(0)} = (\epsilon_1,\dots , \epsilon_{n+i-1}, 0, \epsilon_{n+i+1},
\dots , \epsilon_{2n})$), $1\leq i\leq n$. Clearly, $u \in S, u_i \notin S$
(respectively, $u\notin R, u_i^{(0)} \in R$).

There exists a communication protocol $C_0$ from the family constituting $C$
which gives correct outputs for the input $u$ and for at least of $n/2$ inputs
among $u_i$ (respectively, $u_i^{(0)}$). Without loss of generality one can
assume that the outputs are correct for all $u_i, 1\leq i\leq \lceil n/2 \rceil$
(respectively, for $u_i^{(0)}$). 

Take the path in $C_0$ which follows the input
$u$ and consider the testing polynomials $P_1(q_1,\dots , q_r), \dots , P_N(q_1,\dots , q_r)$
along this path (cf. Section~\ref{protocols}). Denote $P=\prod_{1\leq j\leq N}
P_j(q_1,\dots , q_r)$. We claim that the least term 
$lt(P)=\prod_{1\leq j\leq N} lt(P_j(q_1,\dots , q_r))$ divides on each 
$Z_i, 1\leq i\leq \lceil n/2 \rceil$ (recall that the least term is defined with
respect to the coordinates $(X_1,\dots , X_n,Z_1,\dots , Z_n)$). Otherwise, if
$lt(P)$ does not divide on $Z_i$ then we have

$$sgn(P_j(q_1,\dots , q_r)(u))=sgn (P_j(q_1,\dots , q_r)(u_i)), 1\leq j\leq N$$
  
\noindent
(respectively, 

$$sgn(P_j(q_1,\dots , q_r)(u))=sgn (P_j(q_1,\dots , q_r)(u_i^{(0)}))).$$

\noindent
Hence $C_0$ gives the same output for both inputs $u$ and $u_i$ (respectively,
$u_i^{(0)}$). The obtained contradiction proves the claim.

Thus, the theorem would follow from the next lemma taking into account 
Lemma~\ref{lower}. \bull

\begin{lemma}
If for a certain $k>1$ the product $Z_1\cdots Z_k$ divides $lt(P)$ then for the rank of $n\times n$ matrix 
we have

$$rk \left({\partial ^2 P \over \partial X_i \partial Y_j}\right) \geq k$$
\end{lemma}

\proof Let $lt(P)=p_0 X_1^{m_1} \cdots X_n^{m_n} Z_1^{l_1} \cdots Z_n^{l_n}$ 
where $p_0\in \RR$.
Then the highest term (cf. (\ref{1})) of a non-diagonal entry 
${\partial ^2 P \over \partial X_i \partial Y_j}(u)$ when $i\neq j, 1\leq i,j \leq k$
equals 

$${l_il_j \epsilon_1^{m_1}\cdots \epsilon_n^{m_n}\epsilon_{n+1}^{l_1} \cdots
\epsilon_{2n}^{l_n} \over \epsilon_{n+i} \epsilon_{n+j}}$$

\noindent
The highest  term of a diagonal entry ${\partial ^2 P \over \partial X_i \partial Y_i}(u)$ 
either equals  

$${l_i(l_i-1)\epsilon_1^{m_1}\cdots \epsilon_n^{m_n}\epsilon_{n+1}^{l_1} \cdots
\epsilon_{2n}^{l_n} \over \epsilon_{n+i}^2}$$

\noindent
when $l_i>1$ or is less than 

$${\epsilon_1^{m_1}\cdots \epsilon_n^{m_n}\epsilon_{n+1}^{l_1} \cdots
\epsilon_{2n}^{l_n} \over \epsilon_{n+i}^2}.$$

Denote by $M$ $k\times k$ matrix with the diagonal $(i,i)$-entries $l_i(l_i-1)$
and the non-diagonal $(i,j)$-entries $l_il_j, 1\leq i,j\leq k$. Then
$det(M)= (-1)^{k+1} l_1\cdots l_k (l_1+\cdots +l_k -1) \neq 0$ when $k>1$. 
Therefore, the coefficient of the $k\times k$ minor 

$$det\left({\partial ^2 P \over \partial X_i \partial Y_j}\right)(u)$$

\noindent
where $1\leq i,j\leq k$ at its highest term 

$$(\epsilon_1^{m_1}\cdots \epsilon_n^{m_n})^k 
\epsilon_{n+1}^{kl_1-2}\cdots \epsilon_{2n}^{kl_n-2}$$

\noindent
equals to $det (M)$ and thereby, it does not vanish, which proves the lemma. \bull

\begin{remark}
The same bound as in the theorem holds as well for the (euclidean) closure 
$\overline S$. 
\end{remark}

\begin{corollary}
The probabilistic communication complexity over complex numbers of $R$ is greater
than $n/4$.
\end{corollary}

\proof Having a probabilistic communication protocol $C$ over $\CC$ which recognizes 
$R$, one can convert it into a probabilistic communication protocol $C^{(\RR)}$
over reals which recognizes $R$ at the cost of increasing the complexity at most twice.
For this purpose the first party replaces every polynomial $a(X)$ in $C$ which the
first party calculates by a pair of polynomials $Re(a), Im(a)\in \RR[X]$ in 
$C^{(\RR)}$ where
$a=Re(a)+\sqrt{-1}Im(a)$. The same for the second party. Then for each testing polynomial
$P_j(q_1,\dots , q_r)$ its real and imaginary parts $Re(P_j(q_1,\dots , q_r)),
Im(P_j(q_1,\dots , q_r))$ can be expressed as polynomials over $\RR$ in $Re(q_l),
Im(q_l), 1\leq l\leq r$. Any ramification condition $P_j(q_1,\dots , q_r)=0$ in $C$ we
replace in  $C^{(\RR)}$ by $Re(P_j(q_1,\dots , q_r))=Im(P_j(q_1,\dots , q_r))=0$. To
complete the proof of the corollary we apply Theorem~\ref{main} to $C^{(\RR)}$. \bull

As particular cases consider the problem EMPTINESS: whether the intersection of two
finite sets $\{x_1,\dots ,x_n\} \cap \{y_1,\dots , y_n\} =\emptyset$ is empty? It
corresponds to the arrangement $\cup _{i,j} \{x_i=y_j\}$ (in $\CC^{2n}$ or $\RR^{2n}$).
Another example is the KNAPSACK problem: whether there exist subsets $I_1,I_2\subseteq
\{1,\dots , n\}$ such that $\sum_{i_1\in I_1} x_{i_1} + \sum_{i_2\in I_2} y_{i_2} =0$?
It can be also represented as an arrangement (cf. \cite{G}).

\begin{corollary}
The probabilistic communication complexity of both EMPTINESS and KNAPSACK problems is
greater than $n/4$ over $\CC$ and greater than $n/2$ over $\RR$. 
\end{corollary}

\vspace*{0.4 cm} {\bf Acknowledgements.}
The author is grateful to the Max-Planck Institut fuer Mathematik, Bonn 
where the paper was written, to Farid Ablayev and to Harry Buhrman for 
interesting discussions and to anonymous referees for very detailed comments,
which helped to improve the presentation of the paper.

\end{document}